\documentclass[12pt]{aastex}
\usepackage{emulateapj5}
\usepackage{latexsym}
\usepackage{epsf}

\shorttitle{3-D Logatropic Collapse}

\received{2001 November 24}
\begin{document}

\title{3-D Simulations of the 
Gravitational Collapse of Logatropic Molecular Cloud Cores}
\author{Michael A. Reid, Ralph E. Pudritz, and James 
Wadsley\altaffilmark{1}\altaffiltext{1}{CITA National Fellow}}
\affil{Department of Physics and Astronomy, McMaster University, 
Hamilton, ON, L8S 4M1, Canada}
\email{reidma@physics.mcmaster.ca, pudritz@physics.mcmaster.ca, 
wadsley@physics.mcmaster.ca}

\begin{abstract}
	We present the results of fully 3-D hydrodynamic simulations of
the gravitational collapse of isolated, turbulent molecular cloud cores. 
Starting from initial states of hydrostatic equilibrium, we follow the
collapse of both singular and nonsingular logatropic cores until the
central protostar has accreted $> 90\%$ of the total available mass.  We
find that, in the collapse of a singular core with access to a finite mass
reservoir, the mass of the central protostar increases as $M_{\rm acc}
\propto t^{4}$ until it has accreted $\sim 35\%$ of the total available
mass.  For nonsingular cores of fiducial masses 1, 2.5, and 5 M$_{\odot}$,
we find that protostellar accretion proceeds slowly prior to the formation
of a singular density profile.  Immediately thereafter, the accretion rate
in each case increases to $\sim 10^{-6}$ M$_{\odot}$ yr$^{-1}$, for cores
with central temperature $T_{c}= 10$ K and truncation pressure $P_{s} = 1.3
\times 10^{5} k_{B}$ cm$^{-3}$ K.  It remains at that level until half the
available mass has been accreted.  After this point, the accretion rate
falls steadily as the remaining material is accreted onto the growing
protostellar core.  We suggest that this general behaviour of the
protostellar accretion rate may be indicative of evolution from the
Class~0 to the Class~I protostellar phase. 

\end{abstract}


\section{Introduction}

	A crucial prerequisite for any comprehensive theory of isolated
star formation is an understanding of the gravitational collapse of
molecular cloud cores\footnote{To minimize ambiguity, we will use the
terms ``cloud'', ``clump'', and ``core'' in the following hierarchical
manner:  molecular \emph{clouds} contain dense condensations called
\emph{clumps}, in which entire star clusters may form; clumps, in turn,
contain \emph{cores}, from which single or binary stars may form.  The
centre of a core shall be denoted herein with the terms ``central object''
or ``central protostar''.}.  \citet{Lar69} and \citet{Pen69} pioneered the
study of the collapse of the isothermal sphere from both analytical and
numerical perspectives.  Subsequently, \citet{Shu77} defined the
long-standing paradigm for isolated low mass star formation---the
``inside-out'' or ``expansion wave'' collapse---with his elegant
self-similar solution for the collapse of a singular isothermal sphere
(SIS). 

	A common feature of all three of the aforementioned studies, and
most star-formation studies which have followed, is their adoption of the
isothermal equation of state (EOS).  Recent evidence supports the
assertion that some observed molecular cores are well-fit by isothermal
models.  For example, \citet{Alves} demonstrated convincingly that the
density distribution of the dark cloud Barnard 68 is very well fit by a
critically stable Bonnor-Ebert sphere \citep{B56,E55}. 

	There is, however, mounting evidence which indicates that more
massive molecular cores possess properties which cannot be explained by
the isothermal EOS.  Surveys of prestellar cores in a variety of
star-forming complexes show that massive cores have strong nonthermal
linewidths which are greater than their thermal linewidths \citep{CM95}.  
Furthermore, it has been shown by \citet{FM92} that even low mass 
cores have significant nonthermal linewidths. 

	The SIS model for molecular cloud core collapse predicts a
time-invariant rate of accretion onto the central protostar \citep{Shu77},
which gives rise to a large associated spread between the formation times
of high and low mass stars.  In the high-mass star formation regime, this
is problematic: the constant accretion rate predicted by SIS models is too
low to form massive stars on the timescales suggested by observations
(\citeauthor{CM95} \citeyear{CM95}; \citeauthor{MP97} \citeyear{MP97},
MP97 hereafter).  \citet{FC93} and \citet{BB} have shown that it is
possible to obtain variable accretion rates from isothermal collapse
models if, for example, the mass reservoir is finite or there are initial
deviations from the SIS density profile.  It is not clear, however, that
the mass accretion rates generated by these models vary in a way which is
consistent with observations.  The likely conclusion is that isothermal
models only apply to the formation of low mass stars. 

	The SIS collapse model assumes hydrostatic initial conditions; it
has been suggested that such a singular configuration could arise through
the subsonic collapse of a Bonnor-Ebert sphere \citep{SAL87}.  Simulations
of the collapse of critical Bonnor-Ebert spheres, by contrast, indicate
that, as the density achieves a singular form (i.e. when $\rho(r)$ takes
on an $r^{-2}$ profile), 44\% of the mass is infalling at a few times the
sound speed (\citeauthor{FC93} \citeyear{FC93}; FC93 hereafter).  Such
high infall velocities have not yet been detected in collapsing cores, but
whether this is because they do not exist, or because their detection is
beyond the limits of current observational techniques is not yet known
(see FC93 for discussion). 

	One proposed alternative to the isothermal EOS is the
\emph{logatropic} equation of state, so named for its logarithmic
dependence of pressure on density.  First proposed by \citet{LS89}, the
logatrope is an empirically-motivated EOS which attempts to account for
the presence of nonthermal or turbulent pressure within molecular cores.  
In the so-called ``mixed'' form of \citet{LS89}, the logatrope was
essentially just the isothermal equation of state with a correction added
to account for turbulent support.  \citet{MP96} (MP96 hereafter) proposed
the ``pure'' form of the logatrope, which eliminated the linear
(isothermal) dependence of pressure on density, leaving only the
logarithmic dependence.  In this form, the logatrope can describe the
properties of both low and high mass cores (we refer here primarily to the
properties of molecular cores as determined by spatially unresolved
measurements, such as single measurements of the thermal and nonthermal
components of a core's velocity dispersion).  While nonthermal support is
certainly important in high mass cores, low mass cores are also
characterized by some degree of nonthermal support. 

	The logatrope is capable of addressing some of the difficulties
with the isothermal models.  For example, it is specifically formulated to
account for the nonthermal linewidths of observed cores.  Also, because
they predict a variable protostellar mass accretion rate ($\dot{M}_{acc}
\propto t^{3}$), self-similar models for the collapse of a singular
logatropic sphere (SLS) give rise to a relatively small spread between the
formation times of high and low mass stars. 

	The general consensus at this time favours models which produce
subsonic molecular cloud core collapse.  Because the logatrope is a softer
EOS than the isothermal EOS (i.e. $P$ depends less than linearly on
$\rho$), it has been suggested that it might allow for a gentler, subsonic
approach to the singular density profile (MP96).  This contrasts with the
strongly supersonic flow found in simulations of the collapse of
Bonnor-Ebert spheres (FC93).  The logatrope has the potential to produce
gentler collapses because it accounts for some measure of
turbulent and magnetic support within a molecular cloud core. 

        There is direct observational evidence that the internal (i.e.
spatially resolved) structure of individual turbulent cores matches the
predictions of logatropic models.  For example, \citet{vdT} examined the
structure of the envelopes around 14 massive young stars and found that
the density profiles of these envelopes have power law distributions, $n
\propto r^{\alpha}$, with indices $\alpha = 1.0-1.5$.  This is consistent
with the logatropic model, which predicts $\alpha = 1.0$ for cores in
equilibrium and $\alpha =1.5$ for collapsing cores.  Further, these values
of $\alpha$ indicate density profiles which are significantly flatter than
the $\alpha = 2.0 \pm 0.3$ commonly found for low mass cores, and are
inconsistent with the value of $\alpha = 2.0$ predicted by the SIS model
for an equilibrium isothermal core. These results indicate that nonthermal
support mechanisms dominate in massive young stellar objects, while
thermal pressure dominates in objects of lower mass.  Similarly,
\citet{Henning} and \citet{Colome} find $\alpha = 0.75-1.5$ for the
envelopes around massive Herbig Ae/Be stars.  Other evidence is provided
by \citet{OLA}, who examined the dust thermal spectra of several hot
molecular cores, which are thought to be the sites of massive star
formation.  They showed that the thermal spectra of these cores were best
fit by models whose envelopes had the density profile of collapsing
logatropic spheres. 

	The equilibria, stability criteria, and self-similar collapse
solutions for pure logatropic spheres have been derived by MP96 and MP97. 
They showed that logatropic spheres have two equilibria: one is a stable
equilibrium with finite central density and a $\rho \propto r^{-1}$
envelope, and the other is an unstable equilibrium with a singular ($\rho
\propto r^{-1}$) density profile throughout.  These equilibria are roughly
analogous to those of the isothermal sphere, namely the hydrostatic
Bonnor-Ebert sphere and the unstable SIS.  MP97 also derived self-similar
collapse solutions for the SLS, similar to those developed by
\citet{Shu77} for the SIS. 

	In this paper, we extend the work of MP97 by performing fully
three-dimensional numerical hydrodynamic simulations of the gravitational
collapse of logatropic spheres, both singular and nonsingular.  Our goal
is to develop a set of numerical collapse solutions for the logatrope
which is analogous to those developed for the Bonnor-Ebert sphere by FC93. 
We present results of simulations which describe the approach to a singular
density profile in the collapse of an initially hydrostatic, nonsingular
logatropic sphere.  Because the collapse of a nonsingular logatrope is not
suited to exploration by the usual self-similar analytical techniques, we
are motivated to use a more flexible numerical approach. 

	In \S \ref{sec:lc}, we review the analytic theory of logatropic 
equilibria and collapses.  Section \ref{sec:nm}
describes our numerical method and discusses our testing procedures. 
 Sections \ref{sec:singcol} and \ref{sec:nscol} present the results of our
simulations of the singular and nonsingular logatropic collapses (scaled
such that the total core mass is 1 M$_{\odot}$), respectively.  Section
\ref{sec:obsmatch} discusses the results of simulations of the collapse of
higher-mass nonsingular cores and relates them to the observed properties
of protostellar cores.  We conclude with discussion and a summary
in sections \ref{sec:disc} and \ref{sec:summ}. 

\section{Logatropic Collapse Theory}
\label{sec:lc}

\subsection{The Logatropic Equation of State}
\label{sec:logeqos}

	The logatrope was proposed by \citet{LS89} as a way to model
the observed nonthermal component of the velocity dispersion in molecular
clouds.  Their approach was to separate the equation of state into thermal
and nonthermal components, where they assumed that the nonthermal
component was turbulent in origin, hence $P_{\rm tot} = P_{\rm therm} +
P_{\rm{turb}}$.  In this formulation, $P_{\rm therm}$ is identified with
the isothermal contribution to the pressure, $P_{\rm therm} = P_{\rm iso} =
c_{s}^{2} \rho$.  The form of $P_{\rm{turb}}$ is chosen to fit the
observed correlation between the nonthermal velocity dispersion and
density in molecular clouds:
\begin{eqnarray}
\frac{dP}{d\rho} &=& (\Delta v_{\rm NT})^{2} \propto \rho^{-1} 
\label{eq:lars}\\
P_{\rm turb} &\propto& \ln{\left(\frac{\rho}{\rho_{\rm{ref}}}\right)}~,
\end{eqnarray}

\noindent where $\rho_{\rm ref}$ is the density at some reference point. 
Equation (\ref{eq:lars}) follows simply from the empirical scalings reported
by \citet{Lar81}, namely that the nonthermal velocity dispersion and
density within molecular cores scale as $\Delta v_{\rm NT} \propto
r^{1/2}$ and $\rho \propto r^{-1}$.  MP96 noted that this
isothermal/logarithmic ``mixed'' equation of state had the property that
$P/\rho$ decreased with radius, contrary to observations.  In its place,
MP96 proposed a ``pure'' form of the logatrope: 
\begin{equation}
\label{eq:eqos1}
    \frac{P}{P_{c}} = 1+ \mbox{A} \ln{\left(\frac{\rho}{\rho_{c}}\right)}~,
\end{equation}

\noindent where the subscript ``$c$'' denotes ``central'' values and
$P_{c} = \sigma_{c}^{2} \rho_{c}$.  Here $\sigma_{c}^{2}=k_{B}T_{c}/\mu
m_{\rm H}$ is the central velocity dispersion and $T_{c}$ is the central
temperature of the molecular core.  To account for the fact that the
linewidths in the centres of molecular clouds are observed to be purely
thermal, equation (\ref{eq:eqos1}) is formulated such that, when $\rho =
\rho_{c}$, it gives $P = P_{c} = \sigma_{c}^{2} \rho_{c}$.  Note that it
is \emph{only} at the precise centre of the molecular cloud core that this
correspondence between the isothermal and logatropic models is
established---there is \emph{no} extended central isothermal region. 

	The constant $A$ in equation (\ref{eq:eqos1}) is an adjustable
parameter, and it is always positive.  The best-fit value of $A$ was
determined by MP96, who fit equation (\ref{eq:eqos1}) to observational
data on both low mass \citep{FM92} and high mass \citep{CM95} GMC cores,
obtaining $A = 0.20 \pm 0.02$.  In this work, we take $A=0.2$. 

	It is the pure logatropic equation of state of MP96 whose
properties we explore in this paper.  Henceforth, equation
(\ref{eq:eqos1}) shall be referred to simply as ``the logatrope'', with
the understanding that no further discussion shall be made of other
logarithmic equations of state.

\subsection{Logatropic Equilibria}
\label{sec:logequil}

	In order to examine the equilibria of logatropic spheres, we begin
with the equations governing spherically-symmetric flow in a
self-gravitating fluid: 
\begin{equation}
\label{eq:motion1}
\frac{\partial u}{\partial t} + u \frac{\partial u}{\partial r} 
+ \frac{\partial \phi}{\partial r} + \frac{1}{\rho}\frac{\partial 
P}{\partial r} = 0
\end{equation}
\begin{equation}
\frac{\partial \rho}{\partial t} + 
\frac{1}{r^{2}}\frac{\partial}{\partial r} (r^{2} \rho u) = 0
\end{equation}
\begin{equation}
\label{eq:poisson}
\frac{1}{r^{2}} \frac{\partial}{\partial r}\left(r^{2}\frac{\partial 
\phi}{\partial r}\right) = 4 \pi G \rho~,
\end{equation}

\noindent where $u(r,t)$ is the fluid velocity at radius $r$ and time $t$, 
$\rho(r,t)$ is the density, $P(r,t)$ is the pressure, and $\phi(r,t)$ is 
the gravitational potential.  

\begin{figure*}
\plotone{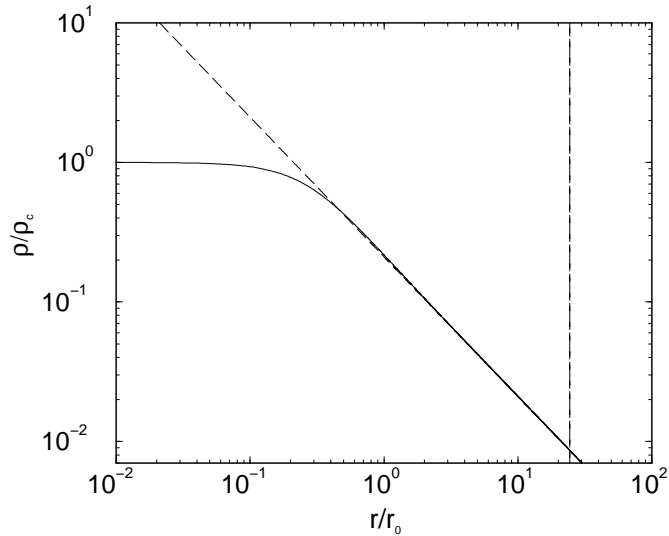} 

\caption{Density profiles for equilibrium logatropic spheres with A=0.20. 
The nonsingular and singular solutions are
represented by the solid and dashed lines, respectively.  The vertical
line represents the truncation radius of a critically stable
cloud (with $R/r_{0} = 24.37$).\label{fig:logdp}}

\end{figure*}

	The equilibria of the logatrope can be found by setting terms in
$u$ and $\partial / \partial t$ to zero in equations
(\ref{eq:motion1})-(\ref{eq:poisson}), inserting the logatropic EOS, and
solving.  One solution that emerges has a singular density profile, of
the form
\begin{equation}
\label{eq:sing}
\frac{\rho}{\rho_{c}} = \sqrt{\frac{2A}{9}}\left(\frac{r_{0}}{r}\right)~.
\end{equation}

\noindent where $r_{0}$, the scale radius of the core, is defined as
$r_{0}^{2} = 9 \sigma_{c}^{2}/4 \pi G \rho_{c}$.  This is the logatropic
analogue of the singular isothermal sphere.  The second solution, which
can be found by numerical integration, represents a pressure-truncated
sphere with a finite central density; this is the logatropic analogue of
the Bonnor-Ebert sphere.  The detailed properties of the two solutions are
developed in MP96, but their density profiles are reproduced here for the
reader's convenience (see Fig. \ref{fig:logdp}).  We note that the
nonsingular profile follows the singular profile closely over a broad
range of radii. 

	MP96 determined the stability criteria for both the 
singular and nonsingular equilibria.  They found that singular logatropes 
of any size are unstable to collapse, but that nonsingular logatropes of 
radius $R$ are stable if
\begin{equation}
\label{eq:stabrad}
R \le R_{\rm{crit}} = r_{0} 
\sqrt{\frac{2A}{9}}\exp\left(\frac{1}{A}-\frac{1}{4}\right)
 \end{equation}
	
\noindent The critical radius of an $A=0.2$ pressure-truncated logatropic 
sphere is $R_{\rm crit}/r_{0}~=~24.37$.

\subsection{Logatropic Collapse}
\label{sec:collapse}

	In \S \ref{sec:singcol}, we compare our SLS simulation results to
the analytic collapse solutions of MP97.  We refer the interested reader
to that paper for a complete treatment of the matter, but we highlight
some of the relevant results here. 

	The SLS collapse solution is qualitatively similar to the SIS
collapse: both are characterized by an expansion wave which begins at the
origin and moves outward through a static medium, causing the material it
passes to collapse inward onto a forming protostar.  In the case of the
SLS, the radius of the expansion wave grows as $r_{ew}=(AP_{c}\pi
G/8)^{1/2} t^{2}$.  The passage of the expansion wave causes material at 
radii $r < r_{ew}$ to collapse,
while material at $r > r_{ew}$ remains stationary.  The expansion wave
reaches the edge of the core at time \begin{equation} t_{\rm surf} =
4\sqrt{2}/\pi \bar{t}_{\rm ff} \simeq 1.8 \bar{t}_{\rm ff}~,
\end{equation} \noindent where $\bar{t}_{\rm ff} =
(3\pi/32\bar{\rho}G)^{1/2}$ is the mean free-fall time in the core. 

	The asymptotic collapse solutions for small radii or late times
are the same for both the SIS and SLS collapse models: $\rho \propto
r^{-3/2}$ and $v \propto -r^{-1/2}$, where $v$ is the mean velocity of the
bulk flow.  One crucial difference between the two models is their
widely-differing mass accretion rates. The SIS collapse is characterized
by a constant rate of accretion onto the central object, $\dot{M} = 0.975
a^{3}/G$, where $a$ is the isothermal sound speed.  In contrast, the SLS 
collapse exhibits a time-varying accretion rate, \begin{equation}
\label{eq:mdotvst} \dot{M}(0,t) = 4(A P_{c} 4 \pi G)^{3/2}
\frac{m_{0}t^{3}}{G}~.  \end{equation}

\noindent Here, $m_{0}$ is a dimensionless parameter---the reduced mass
of the collapsed central object---arising from the self-similar collapse
analysis.  The primary function of $m_{0}$ is to determine the density 
normalization of a given SLS, and thereby to set the mass accretion rate.
For singular logatropes, $m_{0} = 6.67 \times 10^{-4}$, which is
the value used herein.  This scaling of the mass of the protostellar
object with time provides one convenient test of the accuracy of our 
simulations. 

\section{Numerical Methods}
\label{sec:nm}

\subsection{Motivation for Numerical Approach}

        We adopt the numerical approach as a means of extending in time
the self-similar SLS collapse solutions of MP97, and also as a tool for
exploring the collapse of objects which do not lend themselves to the
self-similar approach, such as the nonsingular logatrope.  Further, the
development of a numerical technique is preparatory to future work wherein
we will generalize our approach to study the collapse of rotating,
magnetized logatropic cores.  This purely hydrodynamic study serves to
test our ability to perform accurate 3-D logatropic collapses at feasible
resolutions.  Magnetized, rotating collapses are likely to be
characterized by significant deviations from spherical symmetry, so
their study will be greatly facilitated by the use of 3-D
magnetohydrodynamic (MHD) simulations.

\subsection{The ZEUS-MP Code}

	Our simulations were conducted using ZEUS-MP \citep{SN1,SN2}, a
parallel MHD code with a built-in gravity solver, which we obtained from
the National Center for Supercomputing Applications (NCSA) at the
University of Illinois.  ZEUS-MP uses a finite-difference Eulerian grid
scheme to solve the hydrodynamic equations and a multigrid algorithm to
solve the Poisson equation for self-gravity.  The algorithms are formally 
second order in both time and space.  

	The simulations were run on the McMaster component of SHARC-NET, a
distributed Beowulf cluster shared by several Ontario universities. 
McMaster's component is an AlphaServer~SC with 112 processors and 1 GB of
RAM per processor. 

	Simulations were conducted on uniform Cartesian grids, with
resolutions of both $127^{3}$ and $255^{3}$ cells.  The use of a uniform
Cartesian grid was dictated in part by the ZEUS-MP gravity solver, which
could not be made to perform accurately on other grid types.  

\subsection{Modifications to the Code}

	We made several modifications to the NCSA version of ZEUS-MP.  The
calculation of the pressure term in equation (\ref{eq:motion1}) was
modified to accommodate the logatropic EOS.  ZEUS-MP updates the
velocities in each coordinate direction separately \citep{SN1}.  Thus, in
the $x$-direction, the change in velocity due to the pressure gradient is
typically computed as: 
\begin{equation}
\frac{v_{x,i}(t+\Delta t) - v_{x,i}(t)}{\Delta t} = - \frac{P_{i}(t) - 
P_{i-1}(t)}{\Delta x_{i} (\rho_{i}(t) + \rho_{i-1}(t))/2}
\label{eq:findiff}
\end{equation}

\noindent where $v_{x,i}(t)$ is the velocity in the $x$-direction at grid
point $i$ and time $t$, $P_{i}(t)$ is the pressure, $\rho_{i}(t)$ is the
density, $\Delta t$ is the time step, and $\Delta x_{i}$ is the grid
spacing in the $x$-direction.  ZEUS-MP uses a staggered grid
method, whereby the grid of vector quantities is offset from that of 
scalar quantities by half a grid spacing.  Thus, the velocities in equation
(\ref{eq:findiff}) are computed at a point offset by $\Delta x_{i}/2$ from
the densities and pressures.  Equation (\ref{eq:findiff}) assumes that the
pressures have been calculated previously according to one of the equations
of state built into ZEUS-MP.  For a given equation of state,
however, we can calculate the change in velocity directly, without first
calculating the pressures separately.  Casting the logatropic EOS into
dimensionless units ($P = 1 + A\ln \rho$), we can write the pressure term
as \begin{equation} \frac{1}{\rho}\frac{d P}{d x} = - A \frac{d}{d
x}\left(\frac{1}{\rho}\right)~.  \end{equation}

\noindent  In the finite difference scheme of ZEUS-MP, this gives
\begin{equation}
\frac{v_{x,i}(t+\Delta t)-v_{x,i}(t)}{\Delta t} = - \frac{A}{\Delta x_{i}} 
\left(\frac{1}{\rho_{i}(t)} - \frac{1}{\rho_{i-1}(t)}\right) 
\end{equation}

	Applying this method of calculating pressure forces to the
initial SLS setup gives the analytically expected result in all but the
central $2^{3}$ cells. 

	We also modified the ZEUS-MP gravity solver to improve its
accuracy.  We added the octupole term to the multipole expansion used in
the calculation of the gravitational potential boundary values and we
modified the parallelizaton scheme for this routine.  While the inclusion
of an octupole term in the boundary value routines is not strictly
necessary for a spherically symmetric collapse calculation, we included it
as a precautionary measure intended to more accurately capture any
unexpected deviations from spherical symmetry.  With these changes, the
gravity solver was found to reproduce the expected potential of the
singular logatrope with an error of only 0.1\% in the innermost $2^{3}$
cells, falling rapidly to $3 \times 10^{-4}\%$ at the edge of the sphere.

\subsection{Setting $\rho(r)$ on the Grid}
\label{sec:setrho}

\subsubsection{Density Within the Core}

	The ZEUS-MP gravity solver demands that, when placing a spherical
distribution of mass on a Cartesian grid, scrupulous attention be paid to
the distribution of mass between the sphere and the edge of the
computational grid.  The corners of the grid are especially important
because any significant mass located there will strongly distort the
spherical symmetry of the potential.  To avoid this problem, we surrounded
our spheres with a medium whose density was 1000 times less than that at
the sphere's edge.  This method produced highly spherically symmetric
potentials.  While this density contrast is greater than is expected near
the edges of real molecular cloud cores, tests with lower values showed
that all density contrasts $\gtrsim$ 30 gave similar results.  

	Because the multigrid gravity solver fails to produce accurate
results in the region of a sharp density discontinuity, the jump in
density at the edge of the sphere was smoothed over several cells using a
Gaussian function.  We also found that it was helpful to situate the edge
of the sphere slightly away from the grid boundary.  For a grid
with dimensions $l \times l \times l$, the initial diameter of the
sphere was no more than $0.9l$.

	In order to lay down a singular density profile on a Cartesian
grid, care must be taken in handling the cell containing the singularity. 
Our approach to was to numerically integrate equation (\ref{eq:sing}) to
determine the mass that belonged in the central, singularity-containing
cell, and to divide by the cell's volume to determine its density. 
	
\subsubsection{Density in the External Medium}

	Initially, we elected to treat the low-density outer medium in a
manner similar to that used by FC93.  In that method, the outer 
medium was made isothermal and its sound speed was set to the value 
required to produce the desired truncation pressure at the edge of the 
sphere.  Experiments showed that, while this medium remained at a constant 
pressure for long periods of time during the collapse, it would eventually 
develop unacceptably large fluctuations.  

	A successful treatment of the outer medium was developed by
following the method of \citet{BB}, wherein a constant
pressure---the pressure required to truncate the initial core---is
maintained on the surface of the core throughout collapse.  The boundary
condition is strictly outflow, allowing material to leave the core as
needed but otherwise conserving the mass initially placed within the
boundary.  The velocity of material outside the core is zeroed.  The
effect is to simulate the collapse of a core which has only a finite
reservoir of mass from which to draw.  This is a likely scenario for the
formation of a star in an isolated core, and is also consistent with
observations of clustered star formation in $\rho$~Oph, which indicate
that each accreting object can access only a finite amount of material
\citep{Motte}.

\subsection{A Note on Time Scales}

	In accordance with the convention of \citet{FC93} we define $t=0$
to be the moment during the collapse of the nonsingular logatropic sphere
at which the density profile becomes singular.  Thus, adjustments from the
nonsingular to the singular configurations occur at times $t < 0$. 
The same convention is applied to the singular logatrope: $t=0$ marks the
beginning of the singular collapse.  However, in order to facilitate the
plotting of some of the results of the nonsingular collapse on log-log
scales, we define an alternative time scale, $T=t+t_{\rm sing} \ge 0$
where $t_{\rm sing}$ is the elapsed time between the beginning of the
simulation and the formation of the singularity.  Thus, $T=0$ marks 
\emph{the beginning of the simulation} for both the singular and 
nonsingular cores.  Consequently $T=t$ for the singular core, since it 
begins from a singular density profile (i.e. $t_{\rm sing}$ = 0 for the 
singular core).

	For the purposes of comparison with observations, we have chosen
to scale most of our results using fiducial values of the truncation
pressure, $P_{s}$, and central temperature, $T_{c}$.  In order to
facilitate reinterpretation of our results for other values of $P_{s}$ and
$T_{c}$, we have also provided dimensionless versions of the key results. 
In their dimensionless forms, densities are expressed in terms of the
initial central density, $\rho_{c}$, speeds in terms of the initial
central sound speed, $\sigma_{c}$, radii in terms of $r_{0}$, masses in
terms of the total mass of the core, $M_{\rm tot}$, and times in terms of
the mean free-fall time of the core, $\bar{t}_{\rm ff}$.  All of our
results can be rescaled to other values of $P_{s}$ and $T_{c}$ by 
recalculating each of these quantities using the data in Table 
\ref{tab:sumdat} and the equations in the Appendix of MP96.

\begin{table*}
\begin{center}
\caption{Scaling Parameters \label{tab:sumdat}}
\begin{tabular}{cccc}
\tableline \tableline
$\xi=R/r_{0}$ & $d\Psi/d\xi$ \\ 
\tableline 
1.34 & 0.956 \\
2.21 & 0.959 \\
3.26 & 0.956 \\ \tableline
\end{tabular}
\end{center}
Dimensionless scaling parameters for the three 
nonsingular logatropes.  These values allow for rescaling 
to values of $P_{s}$ and $T_{c}$ other than those used 
herein (see text). 
\end{table*}

\subsection{The Sink Cell}

	The presence of a singularity in the density profile of material
on the grid can lead to two primary problems.  The first is that infalling
material may shock as it encounters the singularity (i.e. the protostellar
object).  The second is that the ever-increasing densities and velocities
in the region around the singularity inevitably cause a crippling trend
toward tiny numerical time steps.  In order to avoid these problems, we
employed a ``sink cell'' approach, modeled after that of \citet{BB}.  The
use of a sink cell effectively isolates the singularity and the details of
the flow around it from the flow on the rest of the grid. 

\begin{figure*}
\plotone{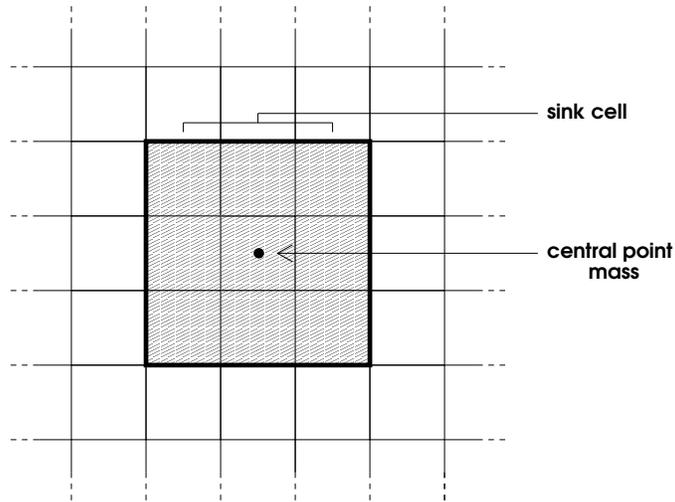} 

\caption{2-D slice through the midplane of the sink cell.  The shaded
cells are inside the sink, which is a 3 $\times$ 3 $\times$ 3 cube of
cells.  All of the shaded cells have the same density, which is equal to
the minimum of the densities of their face-sharing neighbours.  The
location of the point mass is indicated.  The thick line indicates the
border of the sink, where both the gravitational potential gradients and
the pressure gradients are determined by linear extrapolation from the
adjacent cells.\label{fig:sinkgrid}}
\end{figure*}

	In our method, the innermost cube of $3 \times 3 \times 3$ cells
centered on the singularity were blocked off as a collective ``sink
cell''.  Within this region, two types of mass are counted.  The first is
mass which is still on the computational grid and which therefore
undergoes the usual hydrodynamic and gravitational interactions.  The
second type of mass within the sink is that which has passed into a
central point mass; this mass does not interact hydrodynamically with the
rest of the grid, but continues to exert its gravitational influence via a
point mass potential.  The density of grid material within the sink is set
to a spatially uniform value, which is reset at each time step to the
minimum of the densities of the face-sharing neighbour cells (the results
are insensitive to the exact density in this region). We assume that the
mass within the sink can be considered to have condensed onto the central
protostellar object \citep{BB}.  Thus, any mass falling into the sink in
excess of the spatially uniform density therein is subtracted from the
grid and added to the central point mass at the sink's centre.  Figure
\ref{fig:sinkgrid} shows a schematic diagram of our sink cell.

	Because the density of grid material within the sink (i.e. 
excluding the point mass) is not representative of the true density in
that region, we determined the pressure gradients between the cells in the
sink and those immediately outside by linear extrapolation using the
densities in the adjacent two cells.  The gravitational potential
gradients (due to grid matter) around the edge of the sink were determined
in the same way. This was necessary because the multigrid solution for the
potential falters at the edge of the sink, where the density makes the
abrupt change from a singular to a flat-topped profile.  At late times,
the dominant gravitational link between the sink and the rest of the grid
is through the point-mass potential, so the extrapolation of the potential
gradients of material on the grid represents a relatively small
correction. 

\begin{figure*}
\plotone{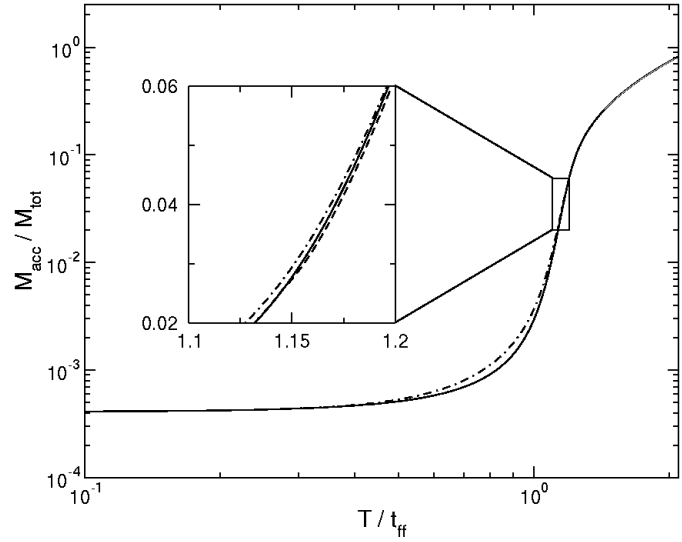} 

\caption{Accreted central mass, $M_{\rm acc}$, in the sink as a function
of time for a 1 M$_{\odot}$ nonsingular collapse.  Mass is expressed as a
fraction of the total mass of the core, and time as a fraction of the
free-fall time.  The solid line represents a simulation with no sink cell. 
The others represent simulations with the sink activated at $T = 0.26
\bar{t}_{\rm ff}$ (dot-dashed line) and $T = 1.14 \bar{t}_{\rm ff}$ (dashed
line).  The behaviour of the three solutions around the time of
singularity formation ($T=1.09 \bar{t}_{\rm ff} = t_{\rm sing}$) is shown in 
the inset. \label{fig:sinkeffect}}

\end{figure*}

	The sink cell can be activated at any time without significant
effect on the collapse.  Figure \ref{fig:sinkeffect} illustrates this
point: it shows the mass accretion profiles for three simulations of the
collapse of a 1 M$_{\odot}$ nonsingular logatropic sphere, differing only
in the presence or absence of a sink cell and its time of activation. The
difference in the mass contained within the central $3^{3}$ cells between
simulations with and without the sink cell never exceeds 5\%, and is $\leq
0.1 \%$ for $t \geq 0.66 \bar{t}_{\rm ff}$.  The effect on the velocities
near the centre is equivalently small.  Both the density and the velocity
far from the centre are essentially unaffected by the use of a sink.  The
small difference visible in Figure \ref{fig:sinkeffect} between the
simulations in which the sink is activated before the singularity forms (T
= 0.26$\bar{t}_{\rm ff}$) and just after (T = 1.14$\bar{t}_{\rm ff}$) is
attributed to the development of supersonic inflow in the region of the
sink just before the formation of the density singularity.  The supersonic
inflow isolates the sink hydrodynamically from the flow on the rest of the
grid, essentially reducing its interaction with the rest of the grid to a
gravitational one.  Thus, if the sink is turned on before this supersonic
inflow develops, its effect on the flow near the centre of the core is
felt more strongly. 

	The use of the sink cell alleviates the problem of shrinking time
steps, extending the time to which the simulation can be followed. 
Additional gains in speed can be made by using lower resolution
simulations.  We use the higher resolution of $255^{3}$ cells in two
situations: for $t < 0$ in the nonsingular collapse, to monitor the
development of the singularity; and for times prior to the departure of
the expansion wave from the core in the singular collapse, for the purpose
of comparison with the analytics.  After these times, however, our primary
goal in both types of simulation is to track the mass accretion history of
the central object, so we switch to the lower resolution of $127^{3}$. 

\begin{figure*}
\plotone{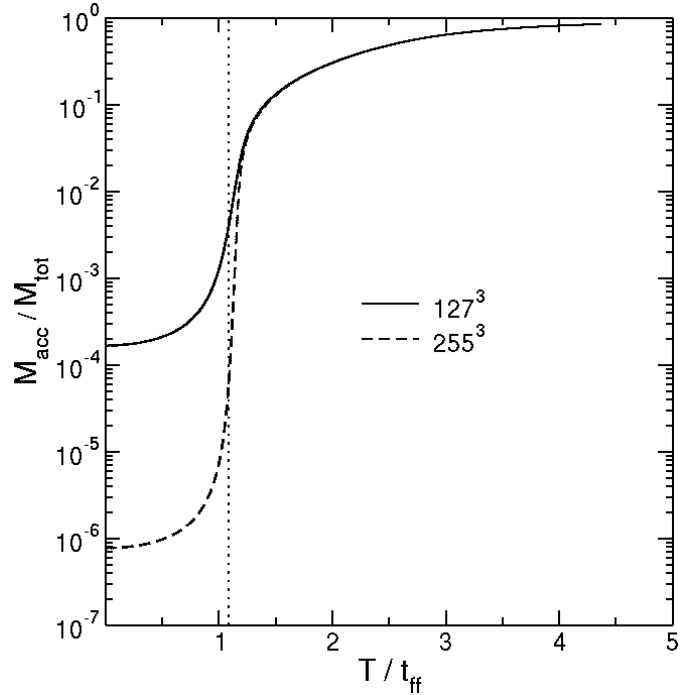} 

\caption{Accreted mass within the sink, $M_{\rm acc}$, as a function of
time for 1 M$_{\odot}$ nonsingular collapses at high ($255^{3}$ cells,
dashed line) and low ($127^{3}$ cells, solid line) resolutions.  The
vertical dotted line indicates the time at which the singular density
profile is achieved.\label{fig:reseff}}

\end{figure*}

	After the singular density profile has been achieved, there is
little difference between the accretion profiles for simulations conducted
at $127^{3}$ and $255^{3}$ cells (see Fig. \ref{fig:reseff}).  When the
simulation begins ($T=0$ in Fig. \ref{fig:sinkeffect}), there is a large
difference in the mass contained within the two sinks simply because the
sink in the higher resolution encompasses a smaller physical volume, and
therefore contains less mass.  After $2.0 \bar{t}_{\rm ff}$,
however, there is only a $3 \times 10^{-3}$\% difference between the sink
masses at the two resolutions.  This is because, by this late time, the
accreted mass is determined almost entirely by the mass of the point mass
(whose dependence on resolution decreases with time).  Hence, to follow
$M_{\rm acc}(t)$ to later times after singularity formation, we are 
justified in switching to the lower resolution to benefit from the improved 
speed. 

\subsection{The Jeans Criterion}
	
	\citet{Truelove} established a criterion for ensuring that 
artificial fragmentation does not arise due to Jeans instability in 
hydrodynamic simulations.  They determined that artificial fragmentation 
will not occur if the ratio of the grid spacing to the Jeans length, 
which they call the Jeans number, $J=\Delta x/\lambda_{J}$, is kept below 
0.25.  We adopt this same implementation of the numerical Jeans criterion, 
with one difference: for the logatrope, the Jeans length becomes
\begin{equation}
\label{eq:logjean}
\lambda_{J} = \sqrt{\frac{\pi}{G \rho} \frac{dP}{d\rho}} = 
\sqrt{\frac{\pi A P_{c}}{G \rho^{2}}}~.
\end{equation}

\noindent  Using this version of $\lambda_{J}$, we check the entire grid 
every few time steps to ensure that $J < 0.25$.

	In all of our simulations, the presence of a central density
singularity inevitably gives rise to violations of the numerical Jeans
condition.  These violations occur only in the immediate vicinity of the
singularity, where the inflow is supersonic.  This supersonic inflow
effectively isolates the Jeans-violating cells from the flow on the rest
of the grid.  For this reason, and because we observe no fragmentation or
significant asymmetry anywhere on the grid, we conclude that these
violations of the numerical Jeans condition near the central cell can be
disregarded. 

\subsection{Testing}

	We tested the code in three ways: we compared test problems to
known results, we examined the quality of equilibria generated, and we
checked that our simulation procedure could reproduce the theoretical
solution for a singular logatrope.  The first two of these tests are
discussed here, and the results of the third can be found in \S
\ref{sec:singcol}.
	
	In order to establish that the code could reproduce known results
when configured for our systems, we ran several of the problems in the
NCSA test suite.  We were able to exactly reproduce published results for
the 1-D Sod shock tube \citep{SN1}.  Further, our simulations of the
gravitational collapse of an initially hydrostatic, uniform-density,
pressure-free sphere on a 3-D Cartesian grid showed excellent agreement
with the analytic results of \citet{Hunter}. 

	Tests showed that the code and our routines could establish and
maintain stable logatropic equilibria.  Recall that we expect nonsingular
cores of radius $R/r_{0} < 24.37$ to exhibit stable equilibria (eq. 
[\ref{eq:stabrad}]).  For a significantly subcritical core ($R/r_{0} =
1.34$ or $M = 1 M_{\odot}$, for our usual choice of central temperature
and truncation pressure) we found that equilibrium could be maintained for
many free-fall times.  After being initialized in hydrostatic equilibrium,
this core oscillated about that equilibrium with velocities never
surpassing $3 \times 10^{-5}$ times the local sound speed for periods of
20 free-fall times (after which we halted the calculation).  Because we
are restricted to using uniform grids, the larger the radius, $R/r_{0}$,
of the simulated sphere, the poorer is our ability to resolve $r_{0}$, and
hence to distinguish numerically between singular (unstable) and
nonsingular (stable) cores.  For this reason, we are unable to obtain
stable equilibria for cores with $R/r_{0} \gtrsim 4.5 $ (i.e. $ \gtrsim 9
M_{\odot}$ for our standard truncation pressure and central temperature),
as these cores behave essentially indistinguishably from singular cores
(even at the higher resolution of $255^{3}$ cells).  Our study of
nonsingular cores is therefore confined to those with truncation radii
$R/r_{0} \leq 3.26$ (corresponding to masses $\le 5 M_{\odot}$), for which
good initial equilibria can be obtained and $r_{0}$ can be properly
resolved over 30 cells or more.  To initiate collapse of the nonsingular
logatrope, we increased the density throughout the core by $5\%$ above
equilibrium values. 

	Establishing and maintaining the unstable equilibrium of the
singular logatrope is considerably more difficult.  As expected, almost
any small perturbation, including those inherent in the integration
algorithms, is sufficient to cause it to collapse.  Our standard
procedure for the singular collapses was to set them up in the best
numerical equilibrium we could achieve, and then initiate the collapse in
a relatively controlled manner by increasing the density of the central
cell by a few percent (the results are insensitive to the exact value of
the density enhancement). 

\subsection{Physical Core Conditions}
\label{sec:physcon}

	Given a value for $A$, the mass of a logatropic sphere is wholly
determined by its radius, $R$, central temperature, $T_{c}$, and
truncation pressure, $P_{s}$.  In all of our simulations, we used $A =
0.2$ in the logatropic EOS.  We take as our ``standard'' central
temperature and truncation pressure the values ($T_{c}$, $P_{s}$) = (10 K,
$1.3 \times 10^5 k_{B}$ cm$^{-3}$ K), chosen to be consistent
with those used by \citet{Shu77} and MP97.  These physical 
conditions are typical of those found in regions of relatively isolated
star formation.  For example, by fitting a Bonnor-Ebert sphere to the
density structure of the isolated dark cloud Barnard 68, \citet{Alves}
derived a surface pressure of $1.8 \times 10^{5} k_{B}$ cm$^{-3}$ K and a
temperature of 16 K. 

	We emphasize that the $T_{c}$ and $P_{s}$ used herein are
\emph{not} representative of conditions in regions of \emph{clustered} star
formation, where the surface pressures can be one or two orders of
magnitude larger (see \S \ref{sec:disc}).  In order to facilitate the
extrapolation of our results to different $P_{s}$ and $T_{c}$ values, we
have included dimensionless versions of the primary results. 

	We have simulated the collapse of nonsingular logatropes truncated
at radii $R/r_{0} = 1.34, 2.21,$ and 3.26.  Subject to our standard
$T_{c}$ and $P_{s}$, these models have masses of 1, 2.5, and 5
M$_{\odot}$ and scale radii of $r_{0} = 0.071, 0.065,$ and 
0.060 pc, respectively.  The higher densities in more massive 
cores lead to greater optical depths, and hence the need to account for
radiative effects. However, as our goal is to study the purely dynamical
aspects of gravitational collapse when radiation pressure is not yet
dominant, our models do not include radiative effects.  Future simulations
of the collapse of higher mass cores will have to account for such
radiative effects. 

	The singular logatrope considered herein is the pure $\rho \propto
r^{-1}$ case of equation (\ref{eq:sing}), corresponding to $C = \sqrt{2}$
and $m_{0} = 6.67 \times 10^{-4}$, in the terminology of MP97.  By
imposing a truncation radius and our standard physical conditions on the
singular logatrope, we scale it such that its total mass is 1 M$_{\odot}$. 

\subsection{From 3-D to 1-D}

	In this paper, we often represent physical quantities (density and
infall speed) as a function of radius within a core.  To convert between
a given quantity represented on a 3-D Cartesian grid and the same quantity
expressed as a function of radius alone, we simply binned the data on the
Cartesian grid into spherical shells, each one pixel thick and centred on
the core's centre.  We then averaged the given quantity over each shell to
plot the quantity as a function of radius alone.

\section{Singular Collapse}
\label{sec:singcol}

	In this section we describe the results of simulations of the
collapse of singular logatropes.  Our objectives in simulating the
singular collapse are (a) to ensure that our method for simulating a
collapse in 3-D accurately reproduces the self-similar solution of
MP97 and (b)~to investigate the accretion history of a singular collapse 
in which the available mass reservoir is finite.

\begin{figure*}
\plotone{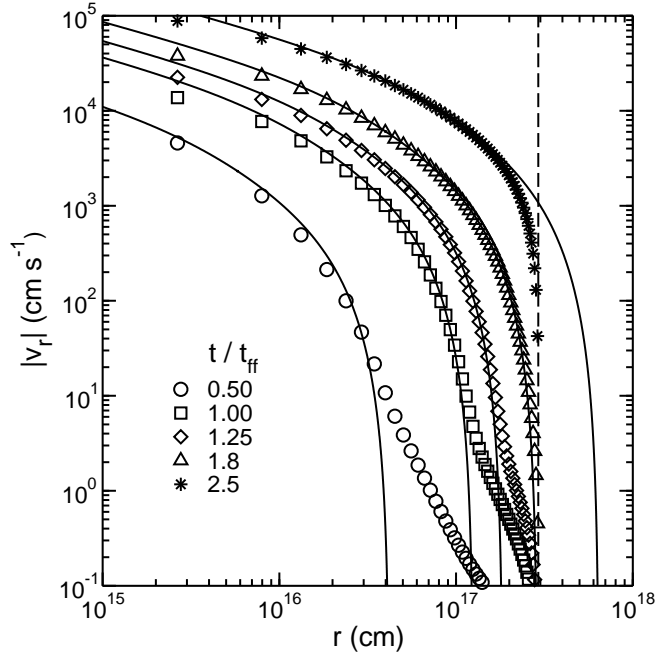} 
\caption{Radial infall speed, $|v_{r}|$, as a function of radius, $r$,
within a singular core, scaled such that $M_{\rm tot} = 1$ M$_{\odot}$.  
Symbols represent the simulation
data, plotted for several different times.  Solid lines are the SLS 
collapse solutions of MP97 at the corresponding times. 
The dashed vertical line marks the truncation radius of the core.
\label{fig:singvel}}
\end{figure*}

	Figure \ref{fig:singvel} shows the time evolution of the radial
infall speed for a singular collapse (scaled such that the total
mass of the core is 1 M$_{\odot}$).  The solid lines in the figure 
are the analytic solutions of MP97 for the SLS collapse.  Each one
represents the infall speed of the gas as a function of radius at 
a given instant. 

\begin{figure*}

\plotone{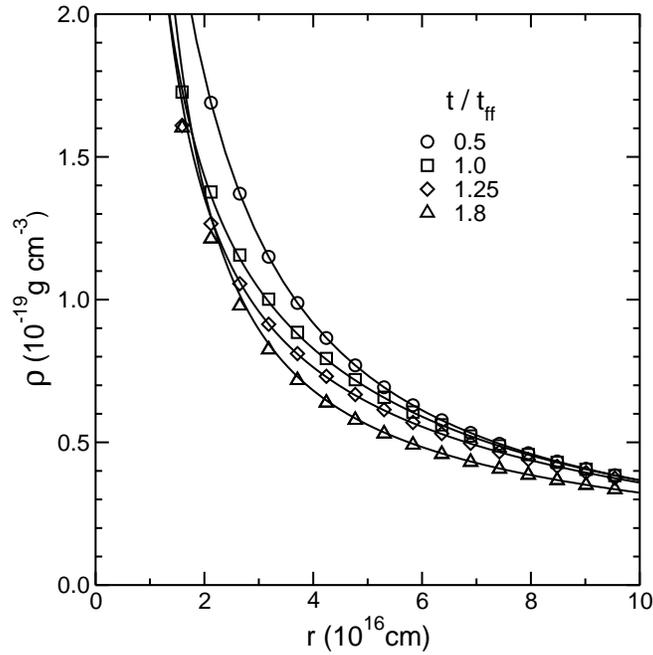} 

\caption{Density, $\rho$, as a function of radius, $r$, within a
collapsing singular core for times prior to $1.8 \bar{t}_{\rm ff}$.  The core
is scaled such that $M_{\rm tot} = 1$ M$_{\odot}$.  The meanings of
symbols and lines are as in Fig. \ref{fig:singvel}.  Only that portion of
the core in which the $\rho(r)$ profiles are distinguishable from each
other is shown. \label{fig:singden}}

\end{figure*}

	In general, the core collapses in the expected inside-out manner
described by MP97.  Due to the initial {25\%} density enhancement applied
to the central cell, the core is initially slightly out of hydrostatic
balance and begins to collapse slowly at all radii.  This effect can be
seen in the figure as the low-speed deviations of the simulation data from
the analytic curves, just past the expansion wave front at each time
interval shown. At early times the expansion wave, which should be
spherical, is poorly resolved on the rectangular grid, so we do not expect
good agreement between simulation and theory.  By 0.5 $\bar{t}_{\rm ff}$,
the expansion wave is easily visible over the small inward motion of the
whole core.  By 1.0 $\bar{t}_{\rm ff}$, the simulation shows good
agreement with the theory.  The density of the collapsing core, shown in
Figure \ref{fig:singden}, also shows good agreement with the analytics of
MP97. 

	Once the expansion wave leaves the core, we can no longer follow 
its behaviour, but we can continue to track the motions of material 
inside the core.  As the 2.5 $\bar{t}_{\rm ff}$ line in Figure 
\ref{fig:singvel} indicates, the collapsing material within the 
truncation radius continues to closely follow the analytic predictions of 
MP97.

\begin{figure*}

\plotone{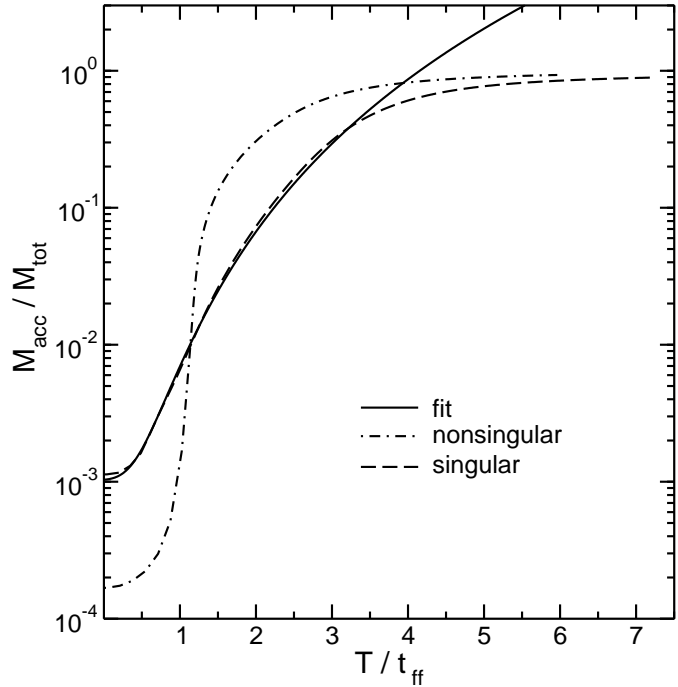}
\caption[Singular vs. nonsingular protostellar 
accretion]{Accreted central mass, $M_{\rm acc}$, vs. time for 
the 1 M$_{\odot}$ singular (dashed line) and nonsingular (dashed-dotted 
line) collapses.  The solid line shows the three-parameter $M_{\rm acc} 
\propto t^{4}$ fit.  Recall that, in both cases, $T=0$ corresponds to the 
moment at which both cores are most nearly singular.\label{fig:sinkacc}} 

\end{figure*}

	Equation (\ref{eq:mdotvst}) tells us that the mass of the central
object at $r=0$ in a collapsing SLS should increase as $t^{4}$.  We can
measure an approximately similar quantity in our simulations by recording
the accretion history, $M_{\rm acc}(t)$, of the sink cell.  Figure
\ref{fig:sinkacc} shows a plot of the accretion history of the
1~M$_{\odot}$ singular core.  Also shown is a three-parameter fit to
$M_{\rm acc} = C_{0}(t+C_{1})^{4} +C_{2}$, where the fitted parameters are
the $C_{i}$.  Here, $C_{0}$ is a scaling factor playing the same role as
the constant $(AP_{c}4 \pi G)^{3/2}$ in equation~(\ref{eq:mdotvst}). The
constant $C_{1}$ accounts for the fact that, at $t=0$, the simulation is
already slightly out of equilibrium, due to the applied density
enhancement.  Thus, the equilibrium state is effectively pushed to a time
$-C_{1}$.  For a wide variety of parameters (resolutions, sink cell
activation times, initial density enhancements, etc.) we find that $C_{1}
\sim \bar{t}_{\rm ff}/5$.  The last constant, $C_{2}$, allows for the fact
that some mass is already present within the sink when it is
activated.  The best-fit values of the $C_{i}$ are $(C_{0},C_{1},C_{2}) =
(2.75 \times 10^{-3},0.210,1.03 \times 10^{-3})$. 

        As predicted by MP97, the accreted mass in the SLS collapse
increases as $t^{4}$ prior to $1.8 \bar{t}_{\rm ff}$, at which time the
expansion wave leaves the core.  After this time, the self-similarity of
the solution is broken and we expect the simulation to diverge from the
self-similar SLS collapse solution.  Because our boundary condition is
such that the accreting central object has a finite mass reservoir
from which to draw (see \S \ref{sec:setrho}), we can predict an
eventual slow-down in the accretion rate.  Once the expansion wave leaves
the core, it is no longer setting new material into collapse, so there
is no new flux of inward-moving material to maintain the flow.

\begin{figure*}

\plotone{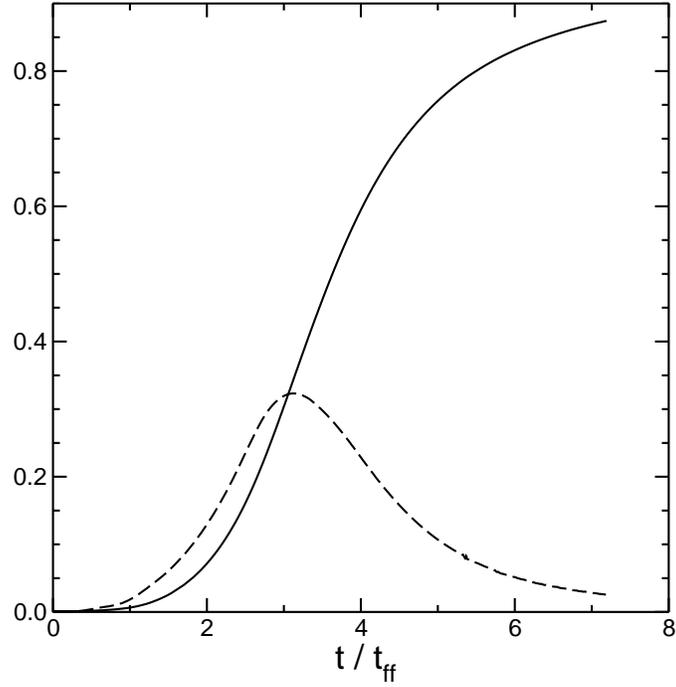}

\caption{Accreted mass, $M_{\rm acc}$, and rate of accretion into the 
sink, $\dot{M}_{\rm acc}$, both vs. time 
for the singular collapse.  The solid line represents $M_{\rm acc}(t)$ as a 
fraction of the total mass and the dashed line represents $\dot{M}_{\rm 
acc}(t)$ in units of $10^{-6}$ M$_{\odot}$ yr$^{-1}$.
\label{fig:mdotvstsing}} 
\end{figure*}

	As shown in Figure \ref{fig:mdotvstsing}, accretion onto the
central object continues as $\dot{M}_{\rm acc} \propto t^{3}$ until $t
\sim 3.1 \bar{t}_{\rm ff}$, at which time 35\% of the total initial mass
of the core is bound up within the central point mass.  From this point
on, the accretion rate declines slowly until we stop the simulation at
$7.2 \bar{t}_{\rm ff}$, when 90\% of the mass of the core has
been accreted.  MP97 predicted that 100\% of the mass in an SLS collapse
would be accreted by $4.3 \bar{t}_{\rm ff}$, but this prediction is based
on the assumption that the self-similar expansion wave solution holds over
the entire collapse.  We have simulated a core with a finite size, so the
simulation ceases to be self-similar once the expansion wave has left the
core.  As our simulations show, the accretion timescale must then
increase, as no new material is being set to collapse by the passage of
the expansion wave.
	
\section{Nonsingular Collapse}
\label{sec:nscol}
	
\begin{figure*}
\plotone{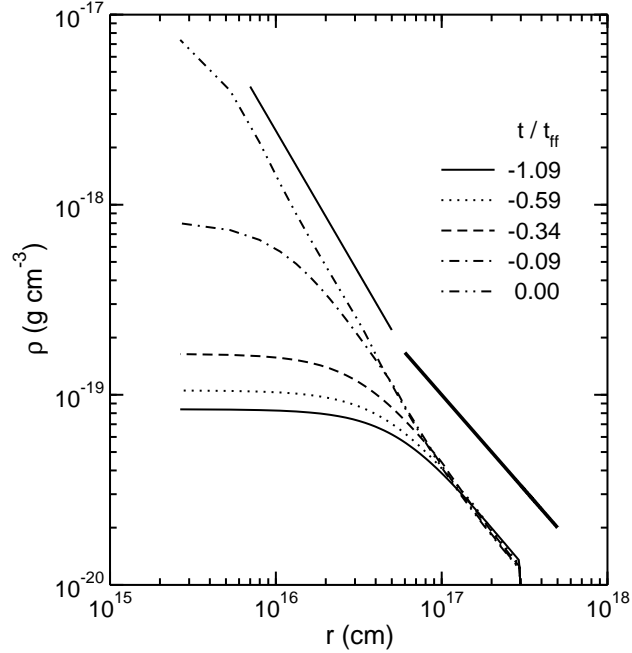}
\caption{Density evolution of the 1 M$_{\odot}$ ($R/r_{0} = 
1.34$) nonsingular core.  
The singular density profile is reached at $t = 0$.  The 
thick straight line indicates $\rho \propto r^{-1}$ and the thinner 
straight line indicates a $\rho \propto r^{-3/2}$ profile. 
The solid curve ($t=-1.09 \bar{t}_{\rm ff}$) represents 
the initial density profile.\label{fig:dvstns}} 
\end{figure*}

\begin{figure*}

\plotone{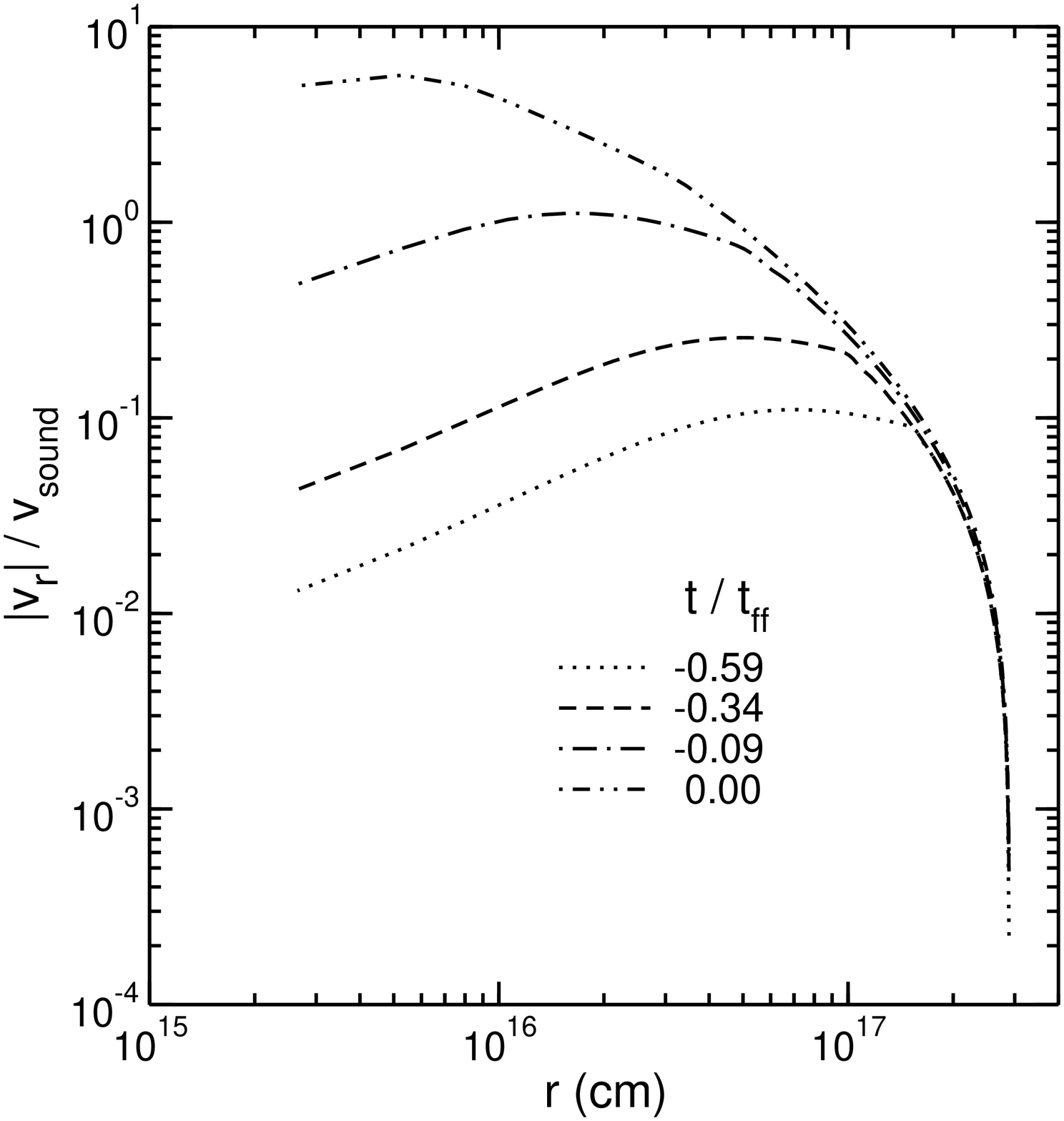}
\caption{Evolution of the radial infall speed of the 1 M$_{\odot}$
($R/r_{0} = 1.34$) nonsingular logatrope preceding the development of the
singular density profile.\label{fig:vvstns}}
\end{figure*}

	The evolution of the 1 M$_{\odot}$ nonsingular core prior to the
formation of the singular density profile is documented in Figures
\ref{fig:dvstns} and \ref{fig:vvstns}.  The nonsingular collapse proceeds
in a manner qualitatively similar to the Bonnor-Ebert collapse of FC93: 
collapse begins at all radii simultaneously (due to the initial applied
overdensity), but the peak in the velocity profile migrates inward slowly. 
The collapse sequence is somewhat different from that seen in the
Bonnor-Ebert sphere.  FC93 found that the inner, initially flat-topped
region of the Bonnor-Ebert sphere collapses to an $r^{-2}$ density
profile, thereby matching the power-law density profile of the rest of the
sphere. We find that the initially flat-topped inner region of the
nonsingular logatropic sphere collapses directly to an approximately
$r^{-3/2}$ density profile.  Thus, at the moment when the inner core
becomes truly singular---the moment we label as $t=0$---the whole core has
a dual power-law density structure: $r^{-3/2}$ for $r \lesssim r_{0}$ and
the initial $r^{-1}$ for $r \gtrsim r_{0}$.  Recall from \S
\ref{sec:collapse} that an $r^{-3/2}$ profile is the form of the density
expected at small radii in the collapse of the \emph{singular} logatropic
sphere.  This would seem to indicate that we are seeing a fairly rapid
transition to an SLS-like collapse.  Note, however, that it is only after
this singular density profile is fully established in the inner core that
we see a marked increase in the central accretion rate (i.e. the
transition from the collapse phase to the accretion phase).  This
transition will be discussed further in \S \ref{sec:obsmatch}. 

	To facilitate comparison between them, Figure \ref{fig:sinkacc}
shows the mass accretion profile for the 1~M$_{\odot}$ nonsingular
collapse in addition to that of the singular collapse.  Prior to the
formation of the singularity, the rate of accretion onto the central mass
increases slowly.  Consequently, very little of the mass is accreted
during this time: by $t=0$, the sink contains only 0.4\% of the total mass
of the core.  Just prior to the emergence of the singular density, the
accretion rate increases markedly.  It peaks just after the singularity
forms, then enters a period of relatively steady accretion until about
half of the total mass has been accreted.  Thereafter, the accretion rate
begins a slow decline which continues for the remainder of the collapse
(see \S \ref{sec:obsmatch} for further discussion).  Table
\ref{tab:nstable} shows a representative set of $M_{\rm acc}$ and
$\dot{M}_{\rm acc}$ values for the nonsingular collapse. 

\begin{table*}
\begin{center}
\caption{Accretion Histories \label{tab:nstable}}
\begin{tabular}{cccc}
\tableline \tableline
$t/\bar{t}_{\rm ff}$ & t & $M_{\rm acc}/M_{\rm tot}$ & $dM_{\rm acc}/dt$ \\
 & ($10^{6}$ yr) &   & ($10^{-6}$ M$_{\odot}$ yr$^{-1}$)\\
\tableline \multicolumn{4}{c}{$\mathbf{1 \mbox{\bf M}_{\odot}}$} \\ \tableline  
-0.504 & -0.244 & $2.36 \times 10^{-4}$ & $7.22 \times 10^{-4}$ \\
0.00 & 0.0 & $4.7 \times 10^{-3}$ & 0.167 \\
0.500 & 0.242 & 0.164 & 0.702 \\
1.00 & 0.484 & 0.340 & 0.760 \\
1.50 & 0.725 & 0.522 & 0.712 \\
2.00 & 0.967 & 0.670 & 0.499 \\
2.50 & 1.21 & 0.768 & 0.325 \\
3.00 & 1.45 & 0.831 & 0.211 \\
3.50 & 1.69 & 0.873 & 0.139 \\
4.00 & 1.93 & 0.901 & 0.0964 \\
4.50 & 2.18 & 0.921 & 0.0702 \\
5.00 & 2.42 & 0.936 & 0.0522 \\
5.43 & 2.62 & 0.945 & 0.0425 \\ \tableline
\multicolumn{4}{c}{$\mathbf{2.5 \mbox{\bf M}_{\odot}}$} \\ \tableline
-0.505 & -0.286 & 2.78 $\times 10^{-4}$ & 1.41 $\times 10^{-3}$ \\
0.00 & 0.00 & 3.80 $\times 10^{-3}$ & 0.238 \\
0.500 & 0.284 & 0.140 & 1.47 \\
1.00 & 0.567 & 0.310 & 1.56 \\
1.50 & 0.850 & 0.488 & 1.50 \\
2.00 & 1.13 & 0.635 & 1.07 \\
2.50 & 1.42 & 0.734 & 0.698 \\
3.00 & 1.70 & 0.798 & 0.451 \\
3.50 & 1.98 & 0.840 & 0.294 \\
4.00 & 2.27 & 0.868 & 0.204 \\
4.50 & 2.55 & 0.888 & 0.148 \\
5.00 & 2.83 & 0.902 & 0.112 \\
5.33 & 3.02 & 0.910 & 0.0940 \\ \tableline
\multicolumn{4}{c}{$\mathbf{5 \mbox{\bf M}_{\odot}}$} \\ \tableline
-0.500 & -0.319 & 4.97 $\times 10^{-5}$ & 3.29 $\times 10^{-4}$ \\
-0.253 & -0.161 & 7.84 $\times 10^{-5}$ & 2.11 $\times 10^{-3}$ \\
0.00 & 0.00 & 1.00 $\times 10^{-3}$ & 0.179 \\
0.250 & 0.159 & 0.0332 & 1.70 \\
0.500 & 0.318 & 0.106 & 2.51 \\
0.750 & 0.478 & 0.186 & 2.58 \\
1.07 & 0.685 & 0.294 & 2.72 \\ \tableline
\end{tabular}
\end{center}
Accretion history of the 1, 2.5, and 5 M$_{\odot}$,
nonsingular, $A=0.2$ logatropic spheres with our standard physical
conditions of $P_{\rm s} = 1.3 \times 10^{5} k_{B}$ K cm$^{-3}$ and
$T_{c}=10$ K.  Recall that $t=0$ marks the time at which the
singular density profile is established.
\end{table*}

\section{Higher Mass Collapses}
\label{sec:obsmatch}

\begin{figure*}
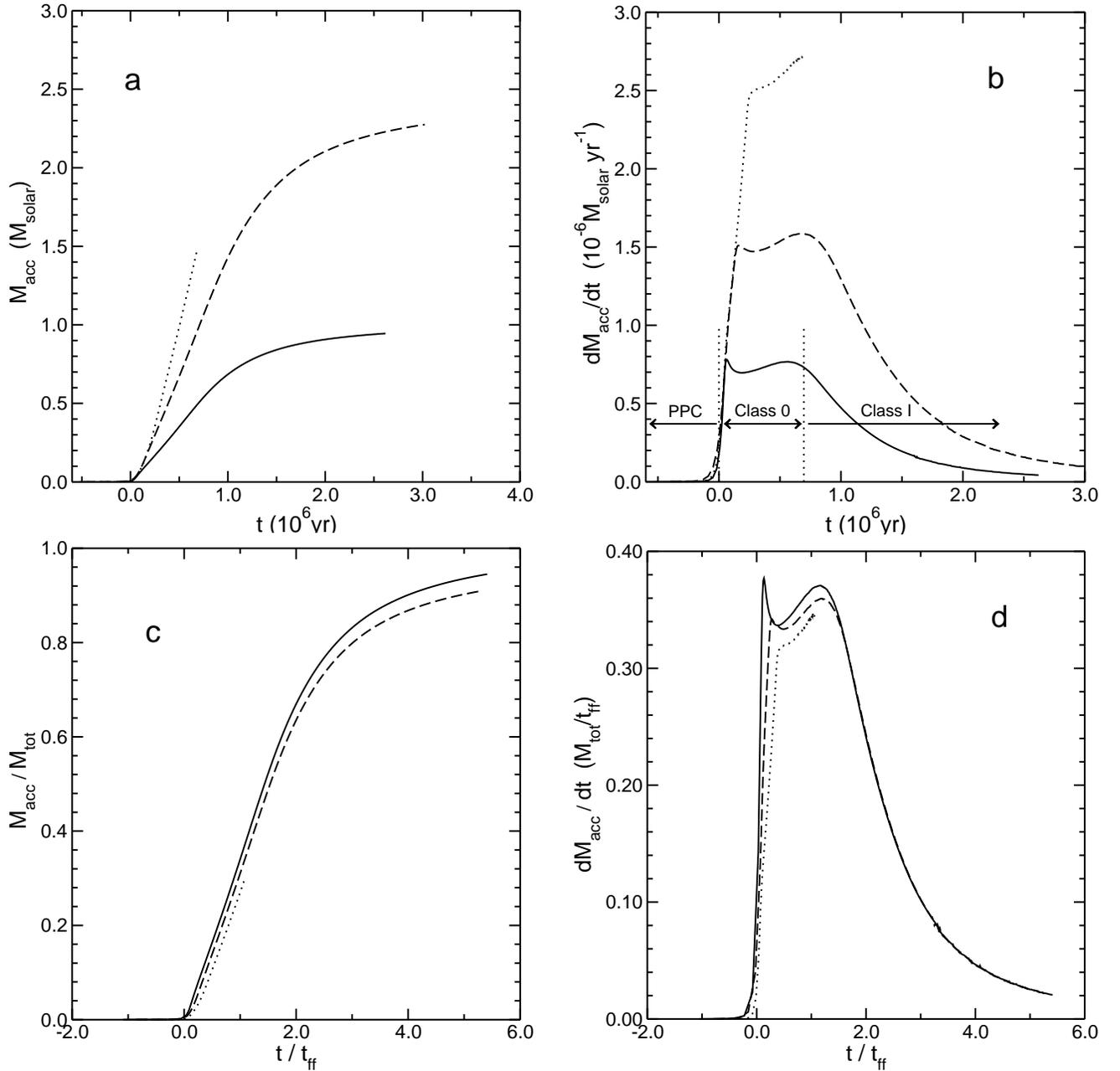

\setlength{\columnwidth}{\textwidth}
\plottwo{f11a.eps}{f11b.eps}
\setlength{\columnwidth}{\textwidth}
\plottwo{f11c.eps}{f11d.eps}
\caption{Time evolution of $M_{\rm acc}(t)$ and $\dot{M}_{\rm acc}(t)$ 
for the collapse of nonsingular logatropic spheres with $R/r_{0} =$ 1.34 
(solid line), 2.21 (dashed line), and 3.26 (dotted line).  Panels (a) 
and (b) show the data scaled according to our standard parameters such 
that the models have total masses of 1, 2.5, and 5 $M_{\odot}$.  Panels 
(c) and (d) show the same data in dimensionless form.  The vertical 
dashed lines in panel (b) indicate the suggested transitions between the 
PPC, Class 0, and Class I stages for the 1 M$_{\odot}$ model (see 
text).\label{fig:mdot}} 
\end{figure*}

	Figure \ref{fig:mdot} shows the time evolution of the accreted
central mass, $M_{\rm acc}(t)$, and central mass accretion rate,
$\dot{M}_{\rm acc}(t)$, for the collapse of nonsingular spheres with
$R/r_{0} = 1.34, 2.21,$ $3.26$.  To situate these results in a physical
context, Figures \ref{fig:mdot}a and \ref{fig:mdot}b are scaled to
according to our standard $P_{s}$ and $T_{c}$, which gives them masses of
1, 2.5, and 5 M$_{\odot}$.  Dimensionless versions of the same data are
supplied in Figures \ref{fig:mdot}c and \ref{fig:mdot}d to facilitate
scaling to other combinations of $P_{s}$ and $T_{c}$.  The $R/r_{0}=3.26$
lines are truncated shortly after the singular density profile is achieved
due to a limitation of the simulation technique.  For this core, switching
to the lower grid resolution to gain the necessary speed improvement would
lead to insufficient resolution of the scale radius, $r_{0}$.  Thus, the 
simulation ends when the time step becomes untenably small at the higher 
resolution.

	According to the singular logatropic collapse theory of MP96, more
massive cores are characterized by higher infall rates, but longer overall
accretion lifetimes.  We find that the nonsingular cores exhibit similar
behaviour.  Contrast this with the SIS model, in which all cores
have the \emph{same} accretion rate, regardless of their total mass. 
Therefore, the accretion lifetimes of higher mass protostars are
proportionately longer than in the logatropic scenario.  As can be seen
in Figure \ref{fig:mdot}b, the central accretion rate (in physical units)
increases with increasing total core mass.  Further, as shown in Figures
\ref{fig:mdot}a and \ref{fig:mdot}c, the 1 M$_{\odot}$ core is the first
to reach M$_{\rm acc}$/M$_{\rm tot}$~=~0.9; that is, the more massive
cores have longer accretion timescales.  The trend seen in Figure
\ref{fig:mdot} then follows logically: in order for the core with the
lowest mass to be the first to accrete 90\% of its mass, it must accrete a
larger fraction of its total mass, per unit time, than the higher mass
cores.  Thus, its accretion rate \emph{expressed in dimensionless units},
must be higher than those of the more massive cores. 

	The accretion histories of all three nonsingular logatropic
models are summarized in Table \ref{tab:nstable}.

	 As in the collapse of the Bonnor-Ebert sphere, in the higher mass
collapses, the central mass of the nonsingular logatropes accretes very
slowly, prior to the formation of the singularity.  In all three cases,
the elapsed time between the initiation of collapse and the formation of
the density singularity is $t_{\rm sing} \simeq 1.44 \tau$, where $\tau =
r_{0}/\sigma_{c}$ is the sound crossing time of the region interior to
$r_{0}$.  This is also just larger than the mean free-fall time of the
region interior to $r_{0}$.  Recall that the initial density profile in
the outer regions ($r>r_{0}$) of the nonsingular logatrope closely follows
the power law profile of the singular logatrope (see Fig.
\ref{fig:logdp}).  Thus, it is reasonable that the time taken for the
global readjustment to power law density behaviour should be of order the
free-fall time of the initially flat-topped region, $r < r_{0}$.  This
information is summarized in Table \ref{tab:mdotab}. 

\begin{table*}
\begin{center}
\caption{Times of Singularity Formation \label{tab:mdotab}}
\begin{tabular}{cccc}
\tableline \tableline
 Mass & $\tau$ & $t_{\rm sing}$ & $t_{\rm sing}$ \\
 (M$_{\odot})$ & ($10^{5}$ yr) & ($\tau$) & ($\bar{t}_{\rm ff}(r_{0}))$\\ 
\tableline
 1 & 4.84 & 1.44 & 1.29\\
 2.5 & 5.67 & 1.45 & 1.35\\
 5.0 & 6.36 & 1.44 & 1.40\\ \tableline
\end{tabular}
\end{center}
Elapsed time between initiation of collapse and singularity 
formation, $t_{\rm sing}$,
for the three nonsingular logatropic spheres subject to our standard 
physical conditions. 
\end{table*}

	For all three masses studied, immediately following the
formation of the singularity, the central mass accretion rate increases
abruptly.  The inflow then moves from the collapse phase into an SLS-like 
accretion phase similar to that seen in Figure \ref{fig:mdotvstsing}. 

	Simulations of the collapse of a critical Bonnor-Ebert sphere
performed by FC93 showed that, at the moment the density profile becomes
singular, 44\% of the mass is in supersonic infall with Mach numbers as
high as $\sim 3.25$.  MP96 predicted that, because the logatropic EOS is
softer than the isothermal EOS, a nonsingular logatropic collapse might
not give rise to such high infall velocities.  Our results are consistent
with this prediction.  We have found that, while highly subcritical
logatropes are characterized by supersonic infall at the moment of
singularity formation, the trend is toward a gentler collapse as we
approach criticality (i.e. as we increase the radius of the core). 

	Table \ref{tab:cftab} lists the percentage of the mass of each
simulated logatropic core which is infalling supersonically at $t=0$.  The
clear trend is that, as the radius of the core increases, both the
fractional amount of mass which is in supersonic infall and the Mach
number of that infall decrease.  Extrapolating the trend, it seems likely
that little or none of the mass of a critical $R/r_{0} =24.37$ logatrope
would be in supersonic infall at $t=0$, in accord with the
prediction of MP96.  Simulations of the collapse of a critical core will 
be needed to confirm this prediction.

\begin{table*}
\begin{center}
\caption{Supersonic Infall Properties \label{tab:cftab}}
\begin{tabular}{ccc}
\tableline \tableline
 $R/r_{0}$ & \% of Mass Infalling & Highest 
$v/v_{\rm sound}$ \\
& Supersonically at $t=0$ & at $t=0$ \\ \tableline
 1.34 & 5.5 & 5.7\\
 2.21 & 2.7 & 4.9\\
 3.26 & 1.3 & 3.2 \\ \tableline
\end{tabular}   
\end{center}
Percentage of
core's total mass which is infalling supersonically at $t=0$, when the
singularity forms.  Note the decreasing trend in both percentage
of mass in supersonic infall and the maximum infall Mach
number with increasing $R/r_{0}$.
\end{table*}

\section{Discussion}
\label{sec:disc}

	As star-forming cores evolve from Class 0 to Class IV, their
protostellar mass accretion rate may vary by $\sim 2$ orders of magnitude
(see \citeauthor{AndrePPIV} \citeyear{AndrePPIV}, and references therein). 
It should be noted that estimates of the accretion rate in young stellar
objects are not made by direct measure, but rather by inference from a
combination of models and other data, as will be discussed shortly.  In
contrast to the observations, the standard SIS model predicts a
time-invariant protostellar mass accretion rate.  \citet{FC93} and
\citet{BB} have obtained variable mass accretion rates for isothermal
collapse models which draw mass from finite reservoirs, but these models
show a tendency to evolve toward the constant mass accretion rate
predicted by the SIS model.  \citet{Henriksen} and \citet{CM95} have
proposed alternative models for protostellar collapse which predict
variable accretion rates.  Like the logatrope, these models are
phenomenologically based, designed to match the observed properties of
star-forming cores (density profiles, nonthermal linewidths, etc.) These
models, however, have many more free parameters than the logatrope, which
has only one: the constant $A$ in equation (\ref{eq:eqos1}).  As shown in
the previous section, the logatrope also predicts a highly variable
accretion rate.  We must ask, then, how closely the logatropic accretion
profile corresponds to available observational data. 

	\citet{Bontemps} surveyed 36 Class~I and 9 Class~0 protostars in
the $\rho$~Oph, Taurus-Auriga, and Perseus star-forming regions and found
that the outflow momentum flux, $F_{\rm CO}$ declines from 
$\sim 10^{-4}$ M$_{\odot}$ km s$^{-1}$yr$^{-1}$ in the Class~0 sources to
$\sim 2 \times 10^{-6}$ M$_{\odot}$ km s$^{-1}$yr$^{-1}$ in the
Class I sources.  Their suggestion was that this decrease in $F_{\rm CO}$
and the accompanying decrease in the mass ejection rate of the
protostellar wind, $\dot{M}_{w}$, were attributable to a decrease in the
protostellar mass accretion rate, $\dot{M}_{\rm acc}$, from $\sim 10^{-5}$
M$_{\odot}$ yr$^{-1}$ to $\sim 2 \times 10^{-7}$ M$_{\odot}$ yr$^{-1}$. 
Clearly the largest of these accretion rates is significantly higher than
those shown in Figure \ref{fig:mdot}, but we note that there is
significant room for movement in both the observations and the scaling of our
simulations. 

	We have scaled our accretion profiles to central temperatures and
truncation pressures which are characteristic of regions of isolated star
formation.  If we instead choose $T_{c}$ and $P_{s}$ in keeping with
observations of regions such as $\rho$ Oph, we can make up the difference
between our $\dot{M}_{\rm acc}$ values and those suggested by
\citet{AndrePPIV}.  \citet{Johnstone} fitted 55 molecular cloud cores in
$\rho$ Oph to Bonnor-Ebert spheres and thereby made estimates of the
surface pressure on each core.  The typical truncation pressures found in
that study lay in the range $P_{s} = 10^{6}-10^{7} k_{B}$ cm$^{-3}$ K. 
These are between 1 and 2 orders of magnitude greater than the fiducial
truncation pressure used herein ($ P_{s} = 1.3 \times 10^{5} k_{B}$
cm$^{3}$ K).  The relationship between the total mass of a nonsingular
logatropic core and its truncation pressure takes the form $M_{\rm tot}
\propto P_{s}^{-1/2}$.  Hence, increasing the truncation pressure by a
factor of 100 would reduce the mass of a given core by a factor of 10. 
This would also bring the mass of the critical $A=0.2$ logatrope down from
92 M$_{\odot}$---typical of an entire molecular cloud \emph{clump}---to
9.2 M$_{\odot}$, which is typical of a molecular cloud \emph{core}
situated within a clump.  Hence, our models would then represent cores of
comparatively low mass.  We would then have to look at logatropic spheres
closer to the critical radius in order to find cores of mass $\sim 1$
M$_{\odot}$, in which case the accretion rates ought to be considerably
higher than those listed in Table \ref{tab:nstable}. 

	The method used by \citet{Johnstone} to extract truncation
pressures was, however, dependent on the assumption that all of the cores
could be well-fit by Bonnor-Ebert spheres, and hence the truncation
pressures extracted cannot necessarily be considered reflective of those
that would be required to truncate logatropic spheres.  The dimensionless
results of Figures \ref{fig:mdot}a and \ref{fig:mdot}b are meant to be
rescalable, should future EOS-independent measures of $P_{s}$ become
available. 

	Disregarding the absolute numbers for a moment (which are, as we
have emphasized, subject to considerable uncertainty), we note that the
\emph{trend} in our logatropic $\dot{M}_{\rm acc}$ bears significant
resemblances to the data.  Estimates place the transition from the Class~0
to the Class~I protostellar phase at the point when the accreted mass is
approximately equal to that remaining in the protostellar envelope,
$M_{\rm acc} \simeq M_{\rm env}$ \citep{AndrePPIV}.  As shown in Table
\ref{tab:mdotab}, the nonsingular logatropic collapse behaves in just this
way---once about half of the mass of the core has been accreted, the
accretion rate, $\dot{M}_{acc}$, begins to fall off.  This behaviour
arises because the accreting central object has only a finite mass
reservoir from which to draw (see discussion in \S~\ref{sec:singcol}).  We
suggest that one interpretation of the mass accretion profiles in Figure
\ref{fig:mdot} is that they describe the transition from Class~0 to
Class~I objects.  If we take the 1 $M_{\odot}$ case as an example, we see
that, at times $t < 0$, there is little
accretion activity, so this may correspond to a very early Class~0 or even
a preprotostellar stage.  From $t \sim 0 $ to $\sim 0.7
\times 10^{6}$ yr, at which time the mass of the protostar and the
remaining envelope are equal, the object undergoes a period of vigorous
accretion in an SLS-like manner.  During this interval, the accretion
rate, while variable, stays relatively constant, never deviating by more
than $6\%$ from the average of $7.4 \times 10^{-7}$ M$_{\odot}$ yr$^{-1}$. 

	If the relationship between the rate at which material is accreted
onto a young protostar and the rate at which mass is ejected through
bipolar jets or outflows were known, it would be possible to extrapolate
the expected outflow rate from this predicted accretion rate.  Unfortunately,
no direct measure of the infall/outflow relationship exists, so we must
resort to making an educated guess based on the available models. 
Following \citet{AndrePPIV} and \citet{PP}, we assume that realistic jet
models give $\dot{M}_{\rm jet}/\dot{M}_{\rm acc} \sim 0.1 - 0.3$. With
this assumption, we extrapolate that accretion at the rate of $7.4 \times
10^{-7}$ M$_{\odot}$ yr$^{-1}$ would be sufficient to power an outflow
with an average $\dot{M}_{\rm jet} \sim 0.74-2.2 \times 10^{-7}$
M$_{\odot}$ yr$^{-1}$.  

	In Figure \ref{fig:mdot}, we have indicated our proposed
evolutionary sequence for the 1 M$_{\odot}$ nonsingular logatropic core
from the preprotostellar to the Class I stage.  The preprotostellar core
(PPC) stage begins with the core in its near-equilibrium hydrostatic
state. The PPC stage ends when the flat-topped central region of the core
has collapsed to a singular $r^{-3/2}$ density profile, just at the onset
of vigorous accretion.  This marks the end of the collapse phase and the
transition to the accretion phase.  The Class 0 stage is characterized by
vigorous accretion throughout.  In accordance with the description of
\citet{AndrePPIV}, we mark the transition from the Class 0 to the Class I
stage when $M_{\rm acc} = M_{\rm env}$.  Once in the proposed Class I
stage, the accretion rates of our simulated cores decline rapidly.  This
decline in the accretion rate correlates well with the observed absence of
strong outflows in Class I cores. 

	One of the appeals of the logatropic model is its relatively small
spread in formation times for stars of a wide range of masses.  The
accretion timescales for the formation of 1-5 M$_{\odot}$ protostars from
nonsingular logatropes are on the order of a few million years, for the
combinations of $R/r_{0}, P_{s}$, and $T_{c}$ used herein.  These
accretion timescales agree well with the observed ages of young star
clusters, such as the Orion Nebular Cluster (ONC), wherein the bulk of the
stars formed within $\sim 10^{6}$ yr \citep{HC2000}.  We anticipate that
the accretion timescale will be shorter in regions of higher pressure, as
in regions of clustered star formation. 

	More sophisticated models of collapsing molecular cloud cores will
include the effects of rotation and magnetic fields.  The former ensures
that most of the material that accretes onto the protostar does so through
a circumstellar accretion disk.  The latter, by contrast, is expected to
cushion the collapse and perhaps to reduce the infall speeds.  Magnetic
fields will also enhance the function of large ``pseudo-disks''
\citep{GS1} through magnetic focusing of infalling material.  It is
important to emphasize that the MP97 models include a measure of
magnetic support, in addition to the turbulent support.  The
logatrope incorporates a mean magnetic field as a virial parameter, but
actual time-dependent simulations employing a full 3-D magnetic field
structure are necessary before one can gain a complete picture of
logatropic collapse. 

\section{Summary}
\label{sec:summ}

	We have conducted 3-D numerical hydrodynamic simulations of the
collapse of singular and nonsingular logatropic spheres, starting from
initial states of hydrostatic equilibrium.  The primary conclusions of this
research are as follows: 

\begin{enumerate}

\item We have extended the solution for the SLS collapse beyond the strict
limit of applicability of the analytic expansion wave solution ($t < 1.8
t_{\rm ff}$).  For an SLS collapse which draws from a finite mass
reservoir, we find that the mass of the central accreting object increases
initially as $t^{4}$, in accordance with the theory of MP97.  This
behaviour continues until $t = 3.1 \bar{t}_{\rm ff}$, when $M_{\rm
acc}/M_{\rm tot} = 0.35$. 

\item We have developed $M_{\rm acc}(t)$ and $\dot{M}_{\rm acc}(t)$
profiles for the collapse of nonsingular logatropic spheres with $R/r_{0}
= 1.34, 2.21$, and $3.26$.  These values of $R/r_{0}$ correspond to 1,
2.5, and 5 M$_{\odot}$ cores, for our choice of fiducial truncation
pressure, $P_{s} = 1.3 \times 10^{5} k_{B}$ cm$^{-3}$ K, and central
temperature, $T_{c} = 10$ K.  We emphasize the accretion results because
they pertain directly to observations of the luminosity of outflows
associated with protostellar cores, and therefore serve as a useful tool
for distinguishing between molecular cloud core collapse models.  The
nonsingular logatropes are all characterized by initially slow accretion,
then a period of intense accretion immediately following the development
of a singular density profile.  The intense accretion continues until
$M_{\rm acc} \simeq 0.5 M_{\rm tot}$, after which the accretion rate
declines gradually until the entire mass of the core is accreted.  For the
three nonsingular collapses, we find that the elapsed time between the
initiation of collapse and the moment of singularity formation (i.e. the
formation of the hydrostatic protostellar object) obeys the approximate
scaling $t_{\rm sing} \simeq 1.44 r_{0}/\sigma_{c} \sim \bar{t}_{\rm
ff}(r_{0})$.  The values of $P_{s}$ and $T_{c}$ used here are typical of 
isolated star formation; rescaling to other pressures and temperatures 
would allow the reinterpretation of these results in terms of clustered 
star formation.

\item We note that the form of $\dot{M}_{\rm acc}(t)$ is qualitatively
suggestive of a transition from the Class~0 to Class~I protostellar
phases.  Improved knowledge of the precise physical conditions ($T_{c}$ 
and $P_{s}$) in the region of specific star-forming cores will be 
required to enable quantitative verification of this hypothesis. 

\item The trend in the accretion profiles is toward dominance of the
envelope-type (i.e. $\rho \propto r^{-1}$) accretion phase with increasing
$R/r_{0}$ (see Fig.  \ref{fig:mdot}).  Simulations of cores with higher
values of $R/r_{0}$ will be needed to confirm this trend. 

\item Our results indicate that, in keeping with expectations based on the
softness of the logatropic EOS (MP96), the adjustment of the nonsingular
logatropic sphere to a singular density profile occurs more gently than in
similar isothermal collapses (FC93).  We showed that, although the collapse
of subcritical nonsingular logatropes gives rise to supersonic
infall at the moment of singularity formation, the fraction of the
infalling mass which is in supersonic infall and the Mach number of the
supersonic flow both decrease as $R/r_{0}$ increases towards the critical
value.  Highly-resolved simulations of the collapse of the critical    
nonsingular logatrope will be required to verify that its approach to the
singular density profile is entirely subsonic.

\end{enumerate}

\acknowledgements

We would like to thank Hugh Couchman for many helpful conversations, Erik
Gregersen for reviewing the manuscript, and Patricia Monger for invaluable
computational support.  M. A. R. was supported in part by a PGS A
scholarship from the Natural Sciences and Engineering Research Council of
Canada.  The work of R. E. P. was supported by the Natural Sciences and
Engineering Research Council of Canada.  J. W. is a CITA National Fellow.

\end{document}